\documentclass{PoS}[11pt]
\pdfoutput=1
\usepackage{graphicx}
\usepackage{graphics}
\usepackage{txfonts}
\usepackage{psfig}
\usepackage{subfig}
\usepackage{sidecap}

\title{Study of the X-ray/gamma source AX J1910.7+0917 and three newly discovered INTEGRAL sources}

\ShortTitle{AX J1910.7+0917 and three new INTEGRAL sources}

\author{\speaker{Lucia Pavan$^a$}, E. Bozzo$^a$, C. Ferrigno$^a$, C. Ricci$^a$, A. Manousakis$^a$, R. Walter$^a$, L. Stella$^b$\\
       $^a$ ISDC - Science data center for Astrophysics,  Universit\'e de Gen\`eve, Switzerland\\
       $^b$ INAF - Osservatorio Astronomico di Roma, Italy\\
       E-mail: \email{lucia.pavan@unige.ch}}

\abstract{ \ax\ is a still unidentified source discovered with ASCA and observed more recently with 
IBIS/ISGRI, mainly noticeable for its rather hard spectrum.\\
We analyzed  all the public available data on this source, and we took advantage of the recent improvements 
performed in the \inte\ data analysis software to fully exploit the IBIS/ISGRI data.
In the data collected from \inte,\ \xmm,\ \chan\ and \asca\ the source is clearly variable. The spectrum can be 
modelled as an absorbed powerlaw ($N_{\rm H}$$\sim$6$\times$10$^{22}$~cm$^{-2}$, $\Gamma$$\simeq$1.5)
with an iron line at $6.4$~keV. 
The present data still do not allow for a unique classification of the source.\\
In the IBIS/ISGRI field of view around \ax,\ we discovered three new sources: 
IGR\,J19173+0747, IGR\,J19294+1327 and  IGR\,J19149+1036, where the latter is positionally coincident with the
\einst\ source 2E\,1912.5+1031.
For the first two sources we report results obtained from follow-up observations carried out with \swift\,/XRT.}

\FullConference{8th INTEGRAL Workshop ``The Restless Gamma-ray Universe''\\
		 September 27-30 2010\\
		 Dublin Castle, Dublin, Ireland}

\def\ax{AX~J1910.7+0917}
\def\inte{{\em INTEGRAL}}
\def\xmm{{\em XMM-Newton}}

\def\chan{{\em Chandra}}
\def\asca{{\em ASCA}}

\def\swift{{\em Swift}}
\def\rosat{{\em ROSAT}}
\def\einst{{\em Einstein}}

\begin{document}

\section{Introduction}
\label{sec:intro}
The wide field of view of the IBIS/ISGRI telescope 
(FOV, 19$^{\circ}$$\times$19$^{\circ}$)\cite{ubertini03} onboard 
\inte\ \cite{winkler03} and its unprecedented sensitivity in the hard X-ray 
domain (17-100~keV), have made this instrument particularly successful 
in the past few years in revealing new high-energy sources.  
To investigate the nature of the still poorly known source \ax,
in this proceeding we use all the publicly available data from \xmm,\ \chan\ and \asca\, 
and we take advantage of the new version of the \inte\ OSA 
software (version 9.0) \cite{courvoisier03} to analyze the \inte\ data.
The analysis of the \inte\ data led also to the discovery of three new hard X-ray 
sources in the IBIS/ISGRI FOV around \ax, independently detected through data analysis also with the    
{\sc bat\_imager} software (A.~Segreto, private communication). 
For the details of the analysis see Pavan et al. 2011 \cite{pavan2011}.\\
\ax\ is a relatively faint and poorly known X-ray source discovered with \asca\ during 
the survey of the Galactic plane. 
The \asca\ spectrum could be fit with an absorbed power-law model
($N_{\rm H} = 2.6^{+1.4}_{-1.0} \times$10$^{22}$~cm$^{-2}$, $\Gamma$=1.1$^{+0.5}_{-0.4}$, with
a flux in the 0.7-10~keV energy range of 2.4$\times$10$^{-12}$~erg/cm$^{-2}$/s).
The source was also detected with \inte\ and reported in the IBIS/ISGRI catalog \cite{bird09}.
No other detections and counterparts in different energy band 
have been reported so far. 
\section{ \ax }
\label{sec:ax} 
\begin{SCfigure}
\centering
\includegraphics[width=0.6\textwidth]{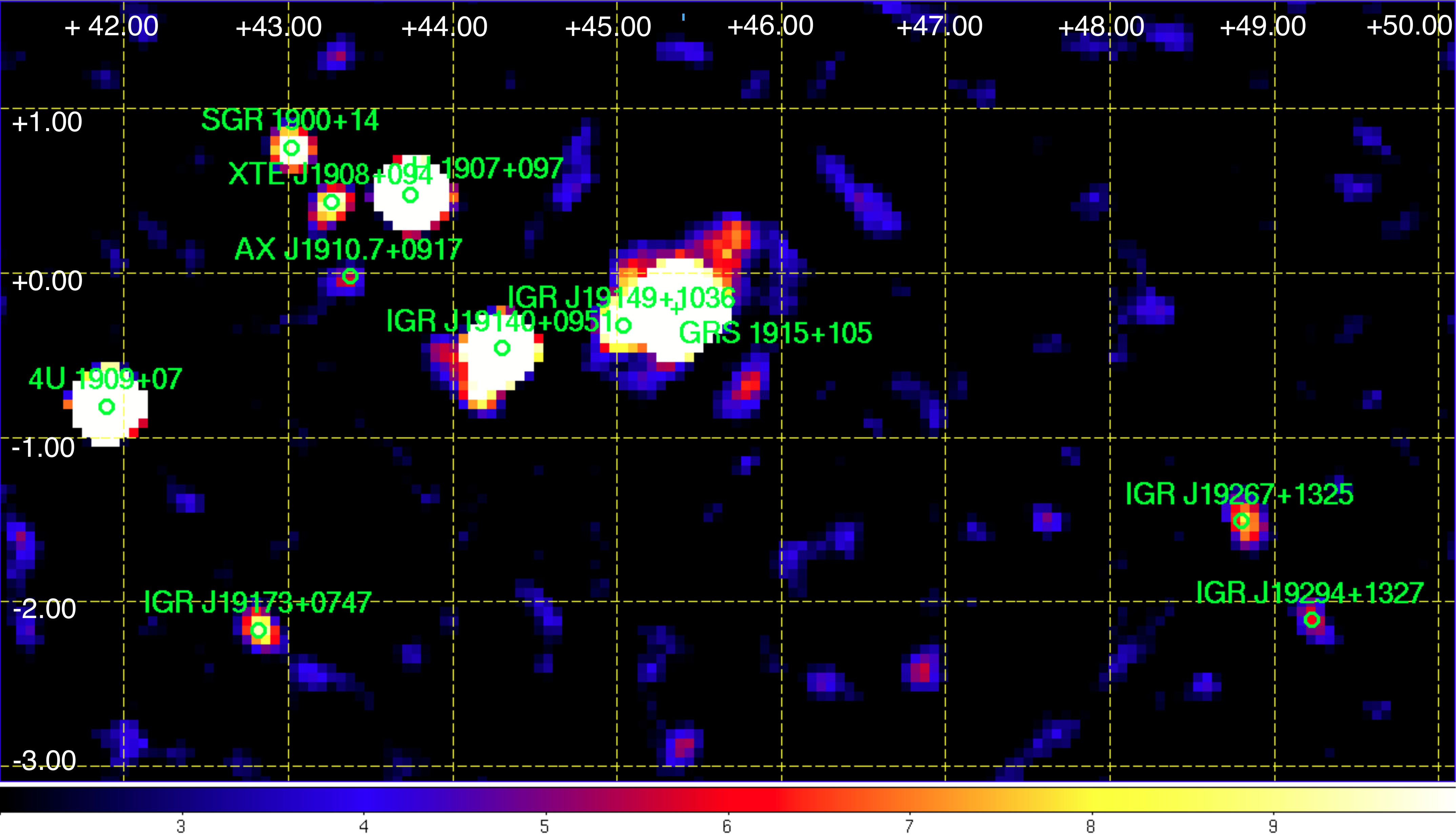}
\caption{IBIS/ISGRI mosaic around \ax\ (17-80~keV, significance map) showing also 
the newly discovered nearby sources. The dashed grids denote galactic coordinates.}
\label{fig:mosaic}
\end{SCfigure}
\begin{figure}
\centering
\subfloat[Light curve. Obs.0084100401]{\label{fig:lc401}\includegraphics[scale=0.216]{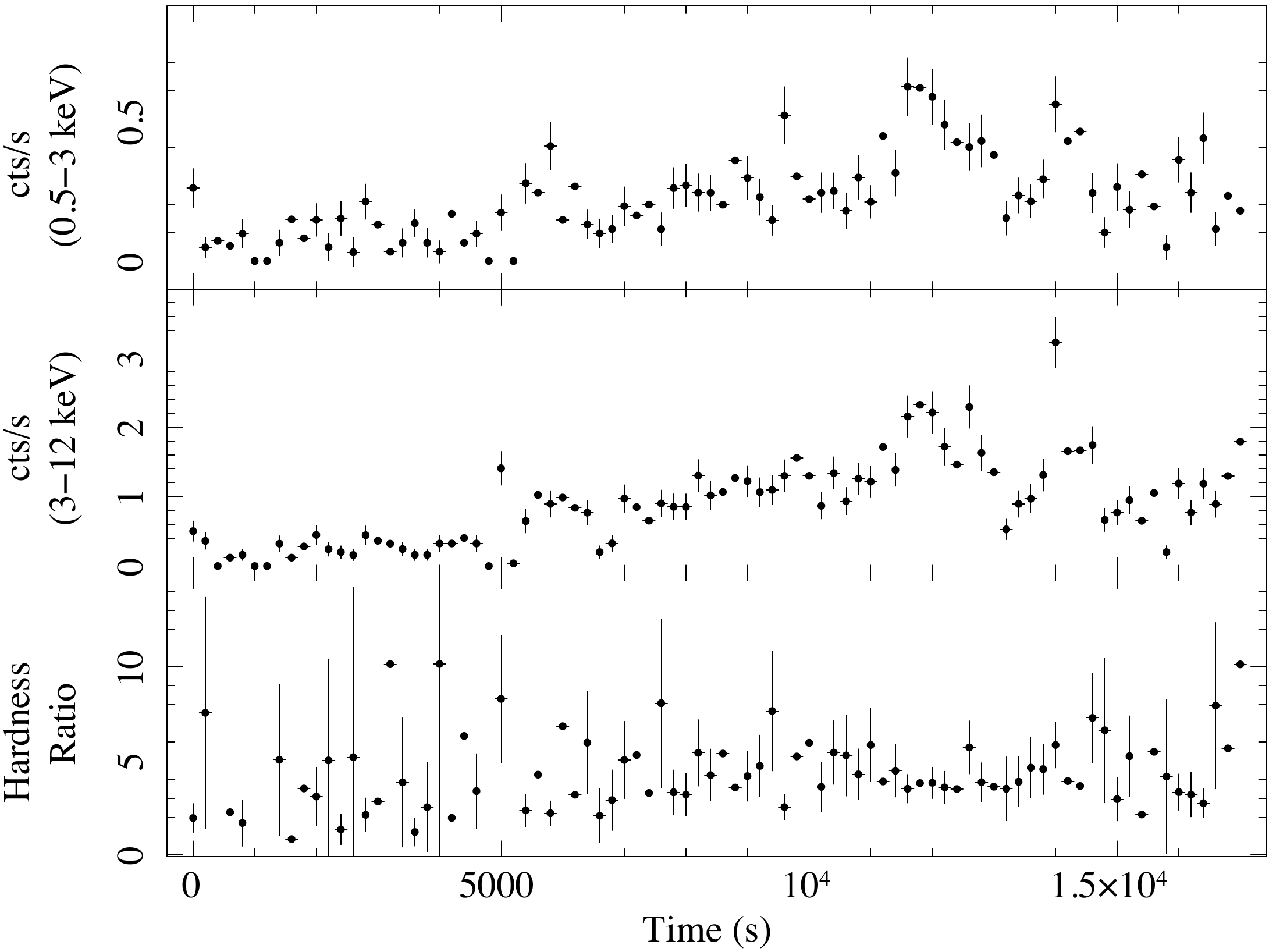}}
\subfloat[Light curve. Obs.0084100501]{\label{fig:lc501}\includegraphics[scale=0.216]{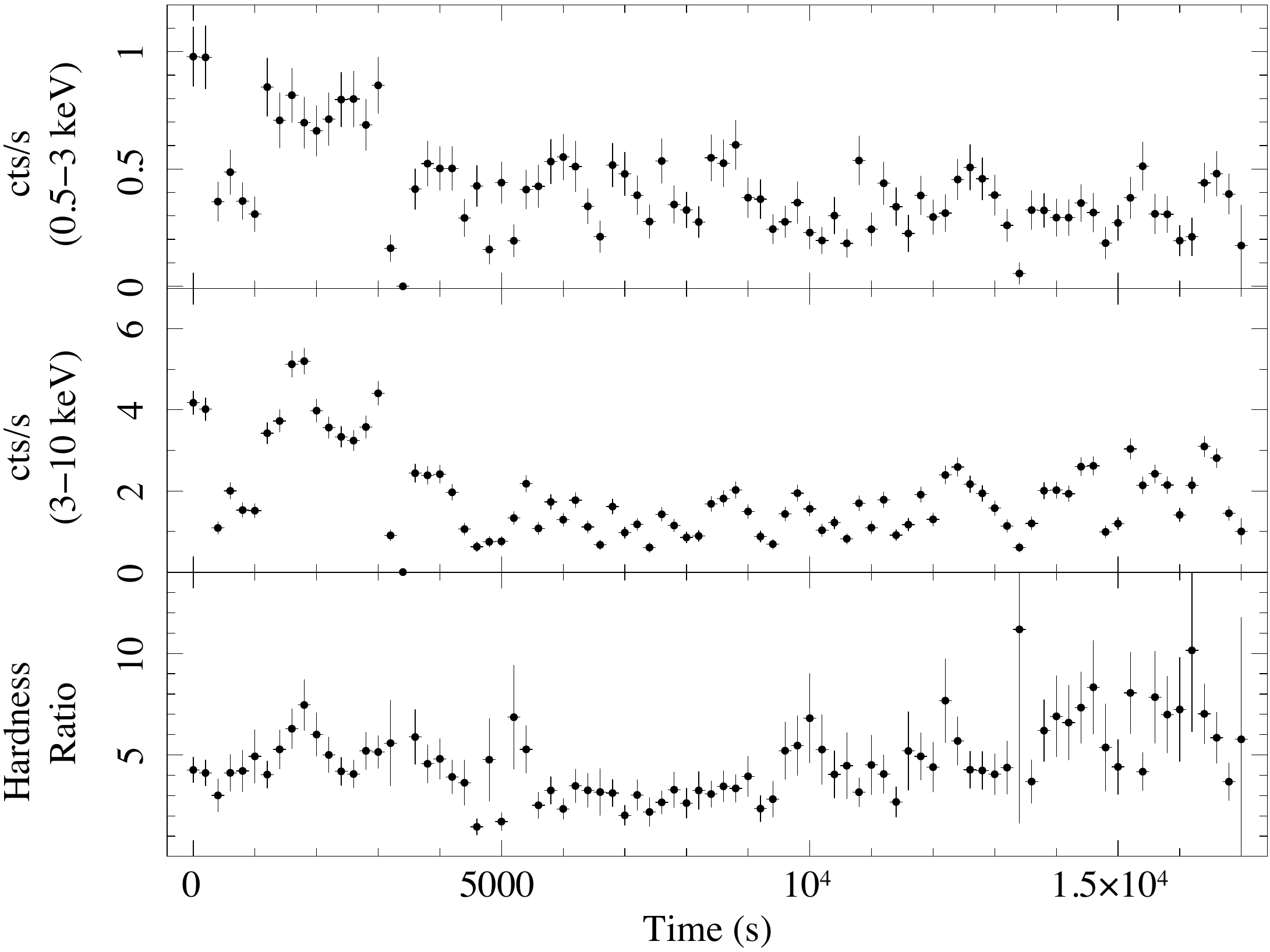}}
\subfloat[Spectrum. Obs.0084100501]{\label{fig:xmmspectrum2}\includegraphics[scale=0.205]{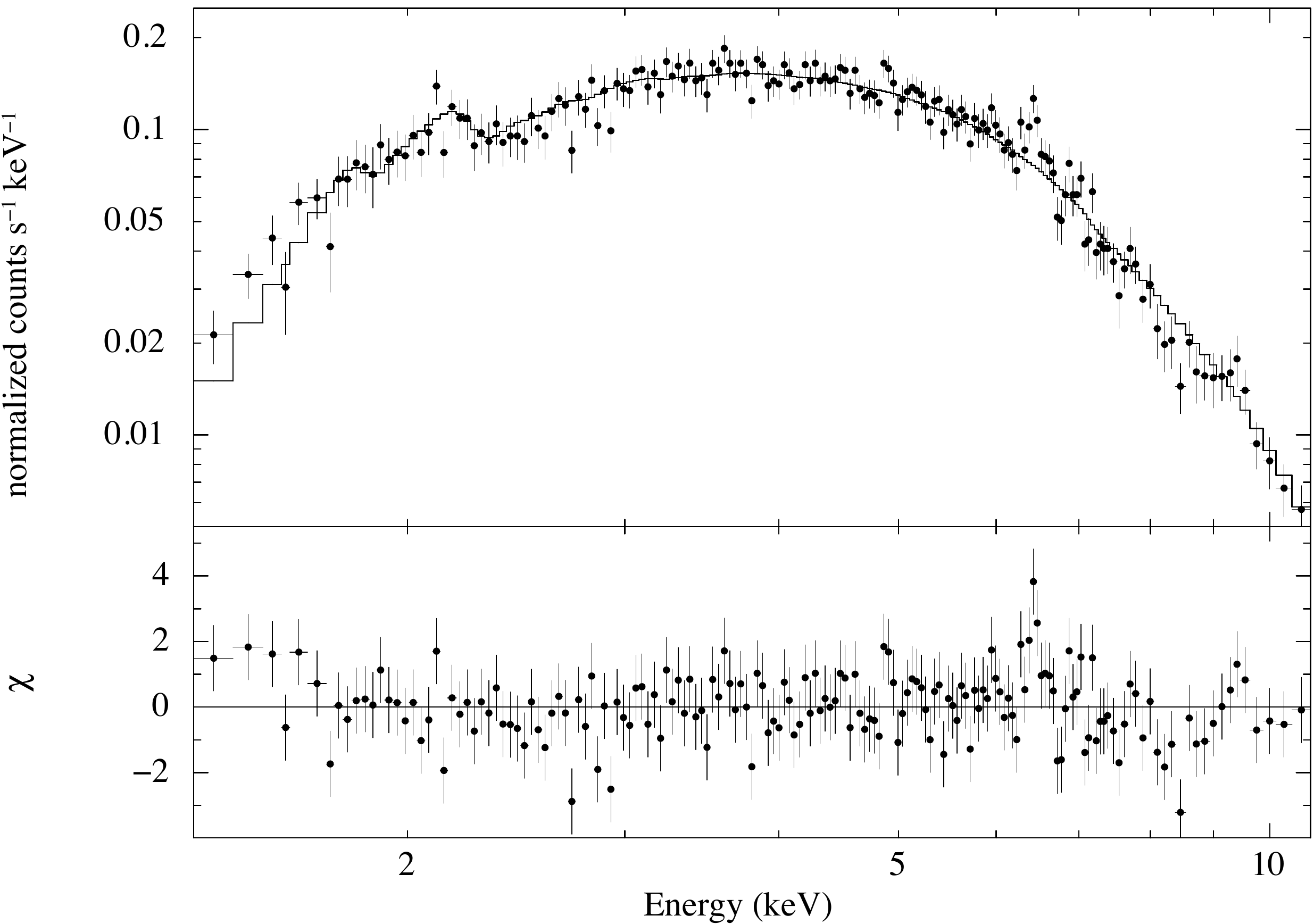}}
\caption{ \xmm\ Epic-pn background subtracted lightcurves and spectrum of \ax. Lightcurves are extracted 
in the two energy bands 0.5-3~keV and 3-12~keV. The hardness ratio is reported 
in the bottom panel of each figure. The time bin is 200 s. The spectrum is shown together with the best fit model 
 (an absorbed power law) and the residuals from the fit, that evidence the presence of a iron line at $\sim$6.4~keV.} 
\label{fig:xmmlcurve}
\end{figure}

\begin{table*}
\begin{center}
\scriptsize
\centering
\caption{ Spectral fit parameters for \ax.\ } 
\begin{tabular}{@{}lllllllll@{}}
\hline
\noalign{\smallskip}
OBS ID & INSTR & DATE$^a$ & EXP$^b$ & $N_{\rm H}$ & $\Gamma$ &  $F_{\rm obs}^{c}$ & $\chi2_{\rm red}$/d.o.f. \\
        &    &           &  (ks)   &  (10$^{22}$~cm$^{-2}$) & & (erg/cm$^{2}$/s) & (C-stat/d.o.f.) \\
\noalign{\smallskip}
\hline
\noalign{\smallskip}
0084100401 & \xmm/Epic-PN & 2004-04-03  & 14.0 & 6.3$^{+0.5}_{-0.4}$ &  1.4$\pm$0.1 & 17.1$^{+1.0}_{-2.1}$ & 1.1/147 \\
\noalign{\smallskip}
0084100501$^d$ & \xmm/Epic-PN & 2004-04-05 & 14.7 & 5.0$\pm$0.3 & 1.28$\pm$0.08 & 24.3$_{-1.7}^{+1.2}$ & 0.9/168 \\
\noalign{\smallskip}
9615$^e$ & \chan/ACIS-S & 2008-05-31 & 1.7 & --- &  --- &   $<$0.4 & --- \\
\noalign{\smallskip}
\hline
\hline
\noalign{\smallskip}
\multicolumn{8}{l}{$a$: Format is YYYY-MM-DD;  $b$: EXP indicates the total exposure time of each observation;} \\
\multicolumn{8}{l}{$c$: Observed flux in the 1-10\,keV energy band in units of 10$^{-12}$;} \\ 
\multicolumn{8}{l}{$d$: This fit includes also a Gaussian line at $\sim$6.4~keV, see text for details;  $e$: 68\%~c.l. upper limit.} \\ 
\end{tabular}
\label{tab:log}
\end{center}
\end{table*}  
%
\label{sec:integral}
We considered all the publicly available \inte\ data 
obtained towards \ax\ until 2009 April 15.
This permitted to achieve an effective exposure time 
on the source of 4.8$\times$10$^2$~ks and 2.7$\times$10$^3$~ks 
for JEM-X (3-23~keV)\cite{lund03} and ISGRI (17-80~keV)\cite{lebrun03} respectively. 
All the data have been analyzed using OSAv9.0
software \cite{courvoisier03}.
The source was not detected in the JEM-X mosaics and
the derived upper limits (1.0-1.7$\times$10$^{-11}$~erg/cm$^2$/s in the 3-7~keV energy band) 
are compatible with the average measured ASCA flux.\\
\ax\ is detected with a significance of 5.8$\sigma$ (in the 17-80~keV band; 
a close view of the ISGRI mosaic is shown in Fig.~\ref{fig:mosaic}).
We derived for the source a count rate of 0.09$\pm$0.02 cts/s, corresponding to
 a flux of 0.31$\pm$0.05~mCrab. 
This is also confirmed by the analysis of the data from \swift\,/BAT (Cusumano, private 
communication), which operates in a similar energy band to that of IBIS/ISGRI. \\
%
\label{sec:xmm}
\ax\ was serendipitously observed in two \xmm\ observations performed 
in 2004 April \cite{miceli06}. 
Due to the low X-ray flux of the source and the relatively short exposure time,
for both observations we report here
only the Epic-pn results for light curves and spectra.
The total effective exposure time is 
of 14.0~ks (14.7~ks) for the Epic-pn in  
observation 0084100401 (0084100501). 
In order to maximize S/N, we extracted source 
lightcurves and spectra by using an elliptical 
region and background lightcurves and spectra from the 
closest source-free region.
In Fig.~\ref{fig:xmmlcurve} we report the Epic-pn lightcurves of the 
source in the 0.5-3~keV and 3-12.0~keV energy bands, extracted from the 
two \xmm\ observations. The hardness ratio, defined as the ratio of the 
count rate in the hard (3-12~keV) to soft (0.5-3~keV) energy band versus time, 
is also shown. A pronounced variability on timescales of hundreds 
of seconds is clearly visible from these lightcurves, but only marginal variations 
in the hardness ratio were measured.
We fit the spectra of both observations with an absorbed power-law model (PL). 
The best fit parameters for both observations are reported in Table~\ref{tab:log}).
In observation 0084100501 the residuals evidenced the presence of an iron line at $\sim$6.4~keV 
(see Fig.~\ref{fig:xmmspectrum2}). 
We thus added a gaussian line to the spectral model used for the fit
obtaining $E_{\rm line}=6.44 \pm 0.03$~keV, with an equivalent width $EW$=0.09$\pm$0.03. 
The normalization of the line was (3.4$\pm$1.0)$\times$10$^{-5}$.\\
Even though no simultaneous \xmm\ and \inte\ observations were available, we fit simultaneosuly the averaged ISGRI and Epic-pn (from observation 0084100501) spectra. This spectrum can still be fit with the same power-law model as in the \xmm\ observations, after introduction of a normalization constant to take into account both the intercalibration between the Epic-pn and ISGRI instruments and the variability of the source. 
The normalization constant obtained from the best-fit is 0.04$\pm$0.02.
This relatively small value indicates that, on average, the X-ray flux 
of the source is much lower than that measured during the \xmm\ observations.\\
%
\label{sec:chandra}  
\ax\ was serendipitously observed in nine \asca\ \cite{tanaka94} observations, performed
 in 1993 and 1999. The spectra extracted in the different observations can be fit
using an absorbed power law model ($\Gamma \sim 1.4$, 
$N_{\rm H} \sim 4.8 \times$10$^{22}$~cm$^{-2}$) with flux ranging from $<0.6$ to $8.2_{-2.7}^{+0.7}\times 10^{12}$ (erg/cm$^{2}$/s) in the 1-10~keV band.\\
The source was also observed by the ACIS telescope on-board \chan\ \cite{garmire03}. 
In the observation ID.~9615, performed on 2008 May 31 and lasted 1.65~ks, 
\ax\ was observed in the FOV of the ACIS-S3 chip but not detected. 
We derived an upper limit on the source 1-10 keV flux of 
4.0$\times$10$^{-13}$~erg/cm$^{2}$/s (assuming a PL model with $\Gamma$=1.4 and 
$N_{\rm H}$=4.8$\times$10$^{22}$~cm$^{-2}$). 
\section{New \inte\ sources} 
\label{sec:newsources} 
In the IBIS/ISGRI FOV around \ax,\ we found three new sources that had previously remained  
undetected. These appeared to be the only excesses 
found independently both in the OSA9.0 mosaic and the mosaic obtained with the {\sc bat\_imager} software 
\cite{segreto10,segreto10b}.
A summary of the properties of the sources is given in Table~\ref{tab:newsources}.
A mosaic containing all the new sources is shown in Fig.~\ref{fig:mosaic}.
\begin{table}
\scriptsize
\centering
\caption{Newly discovered \inte\ sources around \ax.} 
\begin{tabular}{@{}lllllll@{}}
\hline
\noalign{\smallskip}
NAME & RA  & DEC & Err. & DET.$^a$ & Counts$^b$  & Exp.$^c$ \\    
      & (deg) & (deg) & (')  & ($\sigma$)  & & (Ms) \\ 
\noalign{\smallskip}
IGR\,J19173+0747  & 289.349 & 7.785 &  2.1 & 10.0 & 0.15$\pm$0.02  & 3.0 \\
\noalign{\smallskip}
IGR\,J19294+1327 & 292.367 & 13.459 & 3.4 & 6.8 & 0.12$\pm$0.02 & 2.1 \\
\noalign{\smallskip}
IGR\,J19149+1036 &  288.73$^d$&  10.61$^d$& 1.0$^d$& $\sim$20$^d$ &  $\sim$0.3$^d$&  2.6\\
\hline \\
\multicolumn{7}{l}{$a$: Detection significance in the IBIS/ISGRI mosaic (17-80~keV); $b$: The count rates are }\\
\multicolumn{7}{l}{in cts/s estimated from the ISGRI mosaic. In this energy band 1~mCrab=0.28 cts/s; } \\
\multicolumn{7}{l}{$c$: Effective exposure time; $d$: These values are affected by large systematic uncertainties }\\ 
\multicolumn{7}{l}{related to the presence of GRS 1915-105.}
\end{tabular}
\label{tab:newsources} 
\end{table}  
 We report in Table~\ref{tab:newsources} only a first-order approximation for the values 
of IGR\,J19149+1036 as it is relatively close ($\lesssim$20~arcmin) 
to the brighter object GRS\,1915-105 and a precise determination of the degree 
of contamination would require a much more detailed analysis. 
We note, though, that the inferred source position  is coincident with the \einst\ source 2E\,1912.5+1031.\\
For IGR\,J19173+0747 and IGR\,J19294+1327 we obtained follow-up observations (PI L. Stella) in the soft X-ray domain with 
\swift\,/XRT  (0.3-10~keV)\cite{gehrels04}.
We processed all the \swift\,/XRT data by using 
the {\sc xrtpipeline} and the latest calibration files available (caldb v. 20091130). 
Filtering and screening criteria were applied by using {\sc ftools} (Heasoft v.6.9). 
\begin{figure*}
  \centering  
\subfloat[IGR\,J19173+0747]{\label{fig:new3} \includegraphics[width=0.48\textwidth]{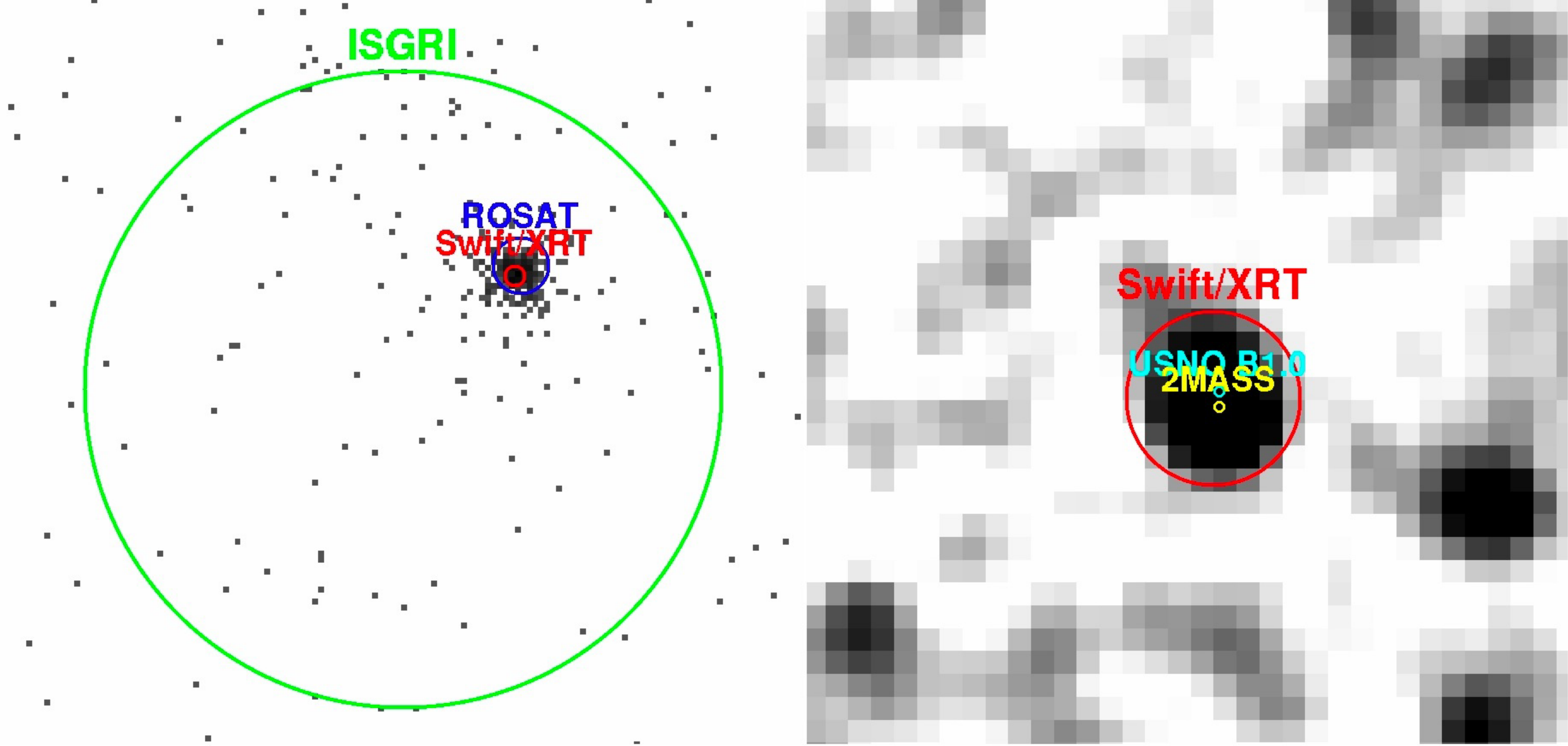}}
\subfloat[IGR\,J19294+1327]{\label{fig:new6}  \includegraphics[width=0.48\textwidth]{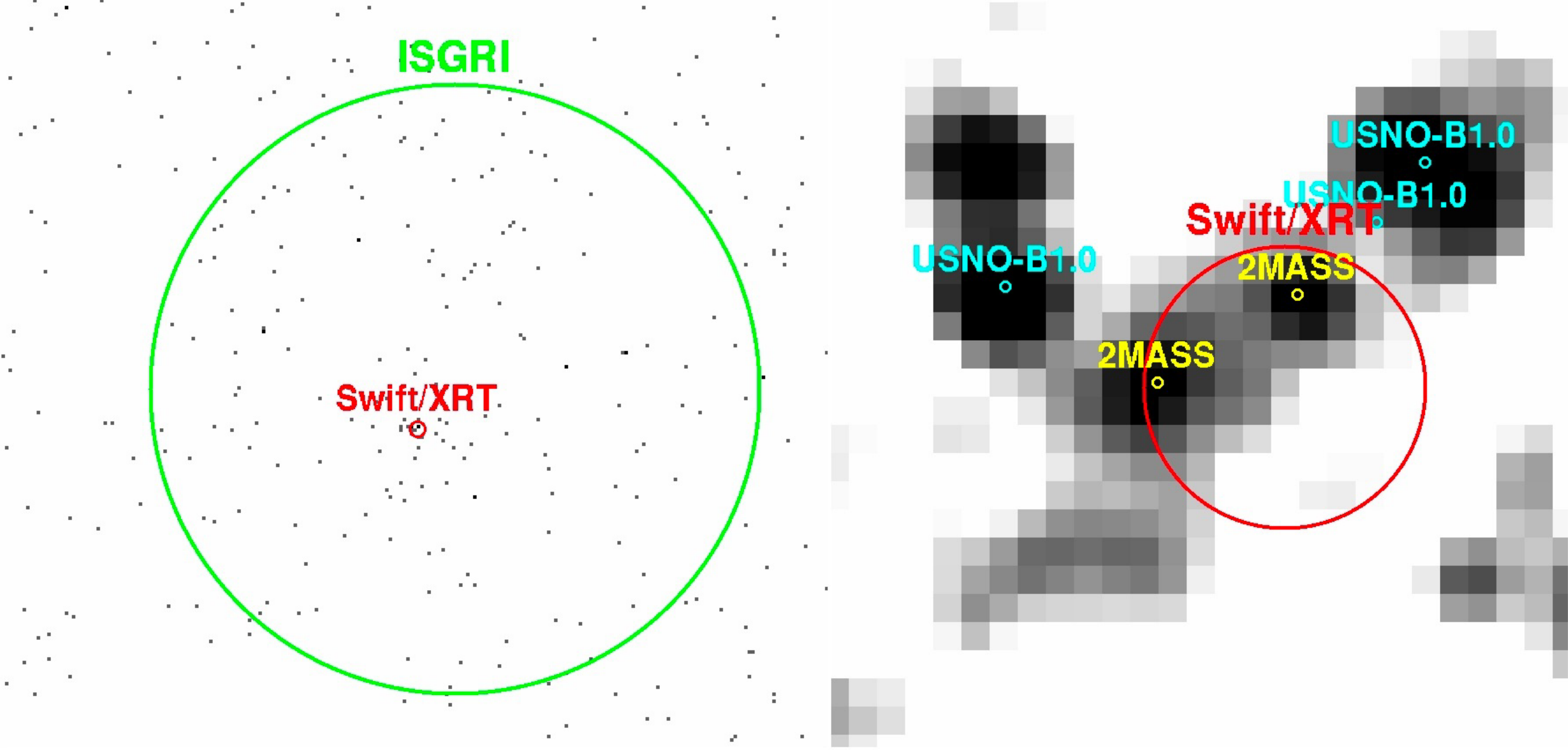}}
  \caption{\swift\,/XRT observations of the newly discovered \inte\ sources. 
 We show on the right of each observation the 2MASS infrared image (J band) together with the \swift\,/XRT 
  error circle and the position of IR and optical counterparts.}  
  \label{fig:swift}
\end{figure*}

IGR\,J19173+0747 was observed by \swift\,/XRT starting on 2010 February 22 at 
08:07:00, for a total exposure time of 6~ks (see fig.~\ref{fig:swift}). 
Inside the \inte\ error circle there is only one soft X-ray 
source. 
The ISGRI spectrum could be well described  
by a power-law model with $\Gamma$=3.3$_{-0.7}^{+0.9}$ ($\chi^2_{\rm red}$/d.o.f.=0.3/5).
while the fit of the \swift\,/XRT spectrum using the same (absorbed)
model gived a PL photon index of $\Gamma$=0.6$\pm$0.2. For the absorption column density
we obtained only an upper limit of $N_{\rm H}$$<$6$\times$10$^{21}$~cm$^{-2}$ (90\% c.l.). 
The corresponding flux is (6$_{-1.5}^{+1.0}$)$\times$10$^{-12}$~erg/cm$^2$/s (0.5-10~keV).
Extrapolating this flux to the 20-40~keV band, would predict a much higher flux ($\sim 2.8 \times 10^{-11}$~erg/cm$^2$/s)
than the one observed with IBIS/ISGRI (5.6$\times$10$^{-12}$~erg/cm$^2$/s). 
This, together with the different photon index derived in the two energy ranges,
suggest a break in the spectrum at energies between 10 and 20~keV or alternatively, 
variability of the source.
We obtained a refined source position at $\alpha_{\rm J2000}$=19$^{\rm h}$17$^{\rm m}$ 20'.8 and 
$\delta_{\rm J2000}$=07$^\circ$47' 51'.1, with an associated uncertainty of 
3.8~arcsec (90\% c.l.). This position is consistent with that of the \rosat\ source 
1RXS\,J191720.6+074755 \cite{voges99}. 
Inside the \swift\,/XRT error circle we found only one possible NIR and optical counterpart.
The NIR counterpart is 2MASS\,J19172078+0747506, characterized by 
J=13.945$\pm$0.031, H=13.520$\pm$0.030, and K=13.311$\pm$0.043. The optical counterpart 
is USNO-B1.0\,0977-0532587 (R1=15.46, B1=16.91, R2=14.99, B2=16.14, I=14.78). 
We queried the FIRST Survey and the NVSS catalogues in search for a radio counterpart, 
but did not find any obvious candidate.

IGR\,J19294+1327 was observed by \swift\,/XRT twice, on 2010 February 22 beginning at 23:59:01 
and on 2010 February 26 beginning at 10:16:01. The total exposure time was 7.4~ks. In the \swift\,/XRT 
FOV only one possible very faint X-ray source is visible within the 
ISGRI error circle (S/N=3.7, see Fig.~\ref{fig:swift}). 
Given the relatively low S/N ratio, other observations are needed to confirm this detection.
The \inte\ spectrum of IGR\,J19294+1327 can be fitted with a powerlaw model with $\Gamma$=2.6$_{-0.7}^{+0.8}$, 
$F_{\rm 20-40~keV}$=6.5$\times$10$^{-12}$~erg/cm$^2$/s ($\chi^2_{\rm red}$/d.o.f.=0.4/4). 

\section{Conclusions}
\label{sec:discussion}

The detailed analysis of \ax\ carried out with \inte,\ \xmm,\ \chan,\ and \asca\
revealed that the source is clearly variable in the soft (1-10~keV) X-ray band (on a relatively short timescale, of hundreds of seconds) and the X-ray spectrum could be well fit using an absorbed power-law model with photon index $\Gamma\sim 1.4$ (consistent with being constant in all the data we analyzed.). We also found that an iron line centered at $\sim$6.4~keV was required  in order to fit the spectrum of the source.
The available X-ray data on \ax\ do not allow for an unambiguous classification of this source. 
Nevertheless, its position relatively close to the Galactic plane, favors the hypothesis of a Galactic source. 
Even though no evidence was found in the \xmm\ and \inte\ data for a coherent periodicity that could be associated 
with the spin period of a neutron star or an orbital period, the width and the centroid of the iron line are compatible with a fluorescence origin and thus suggests that \ax\ is likely part of a binary systems.
The source recalls in particular  the high mass X-ray binaries (HMXB) discovered 
by \inte. Given the variability of the source and the lack of bright and short flares typically seen in SFXTs, \ax\ is possibly a Be X-ray binary.
According to this interpretation, the iron line observed in the \xmm\ spectrum might originate from irradiation of cold iron in the wind of a massive companion, and \xmm\ observation luckily caught the source during an outburst.

Besides carrying out a detailed study of \ax\ in X-rays, we also report the discovery of 
three new hard X-ray sources in the IBIS/ISGRI FOV around \ax.\  
These sources were independently detected with the OSAv9.0 and the 
{\sc bat\_imager} software (A.~Segreto, private communication).
For the two new \inte\ sources IGR\,J19173+0747 and IGR\,J19294+1327, we identified a counterpart in the soft X-ray energy band (0.3-10~keV) thanks to dedicated 
\swift\ observations.
\subsection*{Acknowledgements}
We thanks G. Cusumano for sharing the information 
on the \swift\,/BAT data on \ax,\ A.Segreto for the information on the 
results obtained with {\sc bat\_imager} software, and M. Falanga for useful discussions. 
LS acknowledges financial support from ASI.  

\end{document}